\documentclass[lettersize,journal]{IEEEtran}
\pdfoutput=1
\usepackage{amsmath,amsfonts}
\usepackage{algorithmic}
\usepackage{algorithm}
\usepackage{array}
\usepackage[caption=false,font=normalsize,labelfont=sf,textfont=sf]{subfig}
\usepackage{textcomp}
\usepackage{stfloats}
\usepackage{url}
\usepackage{verbatim}
\usepackage{graphicx}
\usepackage{cite}
\usepackage{color,soul}
\usepackage{multirow}
\DeclareMathOperator*{\argmax}{arg\,max}

\begin{document}

\title{COMIX: Generalized Conflict Management in O-RAN xApps - Architecture, Workflow, and a Power Control case}

\author{Anastasios E. Giannopoulos, Sotirios T. Spantideas, Levis George, \\ Kalafatelis Alexandros, and Panagiotis Trakadas

\thanks{This work was partially supported by the "6G-Cloud" Project, funded by EU HORIZON-JU-SNS-2023 program, under grant agreement No 101139073, and "REACT-6G" project, funded by HORIZON-JU-SNS-2022, 2nd 6G-SANDBOX Open Call, under grant agreement No GA101096328. (Corresponding author: Anastasios Giannopoulos.)}%
\thanks{Anastasios Giannopoulos, Sotirios Spantideas, Levis George, Kalafatelis Alexandros, and Panagiotis Trakadas are with the Research \& Development Department of Four Dot Infinity (FDI), Athens, Greece (Emails: \{angianno,sospanti,glevis,alkalafatelis,ptrakadas\}@fourdotinfinity.com).}%
}
\markboth{{Preprint}}%
{Giannopoulos \MakeLowercase{\textit{et al.}}: COMIX: Generalized Conflict Management in O-RAN xApps - Architecture, Workflow, and a Power Control case}


\maketitle

\begin{abstract}
Open Radio Access Network (O-RAN) is transforming the telecommunications landscape by enabling flexible, intelligent, and multi-vendor networks. Central to its architecture are xApps hosted on the Near-Real-Time RAN Intelligent Controller (Near-RT RIC), which optimize network functions in real time. However, the concurrent operation of multiple xApps with conflicting objectives can lead to suboptimal performance. This paper introduces a generalized Conflict Management scheme for Multi-Channel Power Control in O-RAN xApps (COMIX), designed to detect and resolve conflicts between xApps. To demonstrate COMIX, we focus on two Deep Reinforcement Learning (DRL)-based xApps for power control: one maximizes the data rare across UEs, and the other optimizes system-level energy efficiency. COMIX employs a standardized Conflict Mitigation Framework (CMF) for conflict detection and resolution and leverages the Network Digital Twin (NDT) to evaluate the impact of conflicting actions before applying them to the live network. We validate the framework using a realistic multi-channel power control scenario under various conflict resolution policies, demonstrating its effectiveness in balancing antagonistic objectives. Our results highlight significant network energy savings achieved through the conflict management scheme compared to baseline CMF-free methods.
\end{abstract}

\begin{IEEEkeywords}
6G, conflict detection, conflict management, deep reinforcement learning, energy efficiency, O-RAN, power control, xApp.
\end{IEEEkeywords}

\section{Introduction}

\subsection{Open RAN Overview}\label{subsec:oran}

\IEEEPARstart{T}{he} Open Radio Access Network (O-RAN) architecture represents a transformative approach to the design and operation of radio access networks, emphasizing openness, intelligence, and interoperability \cite{polese2023understanding}. At its core, O-RAN fosters a disaggregated architecture where components such as the Radio Unit (RU), Distributed Unit (DU), and Central Unit (CU) can be provided by different vendors, promoting a multi-vendor ecosystem. This vendor-agnostic framework enables operators to deploy a flexible and cost-efficient network by selecting the preferred components \cite{abdalla2022toward}. Central to this architecture is the RAN Intelligent Controller (RIC), which is divided into the Non-Real-Time RIC (Non-RT RIC) and the Near-Real-Time RIC (Near-RT RIC). These components host specialized applications—rApps on the Non-RT RIC and xApps on the Near-RT RIC—which drive network intelligence and optimization. rApps, with their long timescales, provide strategic guidance, while xApps focus on fine-grained adaptations to network dynamics in real time. By enabling seamless integration of vendor-specific or third-party applications, O-RAN not only accelerates innovation but also enhances adaptability in addressing diverse network scenarios \cite{d2022orchestran}.

The Near-RT RIC plays a pivotal role in achieving intelligent and dynamic control of the RAN through its support for closed-loop optimization \cite{giannopoulos2022supporting}. By operating on timescales ranging from 10 milliseconds to 1 second, the Near-RT RIC manages critical tasks such as resource allocation, interference management, and power control in near real-time. This level of responsiveness enables xApps to directly influence the behavior of RAN components like RUs, DUs, and CUs, ensuring adaptability to rapidly changing network conditions \cite{spantideas2023intelligent}. The Near-RT RIC’s tight control loop complements the slower, policy-driven adjustments of the Non-RT RIC, enabling a hierarchical and time-sensitive optimization framework. This separation of control loops ensures scalability and efficiency while allowing the integration of diverse xApps targeting distinct objectives, such as enhancing data rates or improving energy efficiency.

\subsection{Conflict Management in O-RAN xApps}\label{subsec:conflictmanagement}

The dynamic and decentralized nature of O-RAN introduces significant challenges in ensuring harmonious operation among multiple xApps and/or rApps running concurrently on the RICs \cite{yungaicela2024misconfiguration}. A conflict arises when the actions or objectives of one application negatively impact the performance or objectives of another \cite{adamczyk2023conflict}. These conflicts can degrade the overall efficiency, stability, and fairness of the network, leading to contradicting decisions. For example, unresolved conflicts between xApps managing power control or resource allocation can lead to suboptimal utilization of network resources, increased interference, and inconsistent user experiences. Consequently, robust conflict detection mechanisms are required to identify potential issues proactively, and effective conflict resolution strategies are essential to restore balance and ensure that the system functions optimally without favoring specific applications unjustly. To address this challenge, O-RAN Alliance specifies a Conflict Management Framework (CMF) within the RIC architecture \cite{o2023ran}. The CMF is tasked with detecting, classifying, and resolving conflicts among xApps and rApps to ensure optimal coordination and resource utilization.

\begin{figure}[t]
\centering
\includegraphics[trim={0.4cm 1cm 0.4cm 0.6cm},clip,width=\columnwidth]{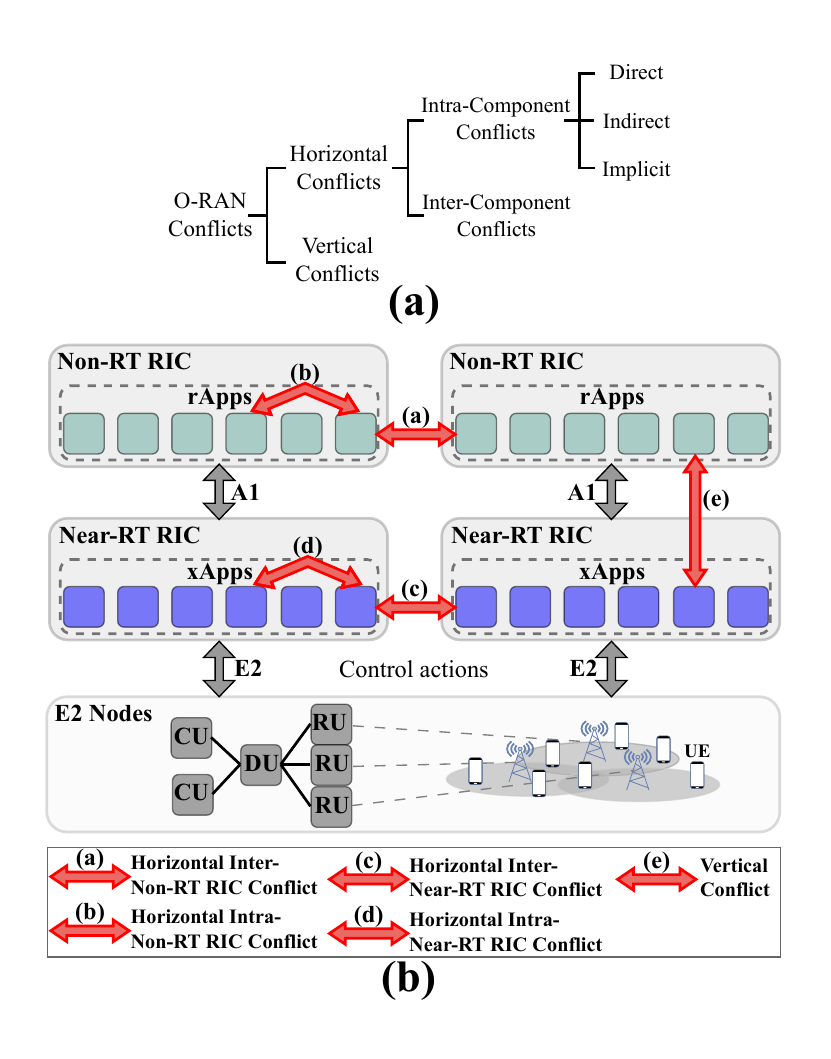}
\caption{Conflicts in O-RAN. \textbf{(a)} Conflict categories in O-RAN. \textbf{(b)} Architectural view of conflict categories.}
\label{fig:fig1}
\end{figure}

Conflicts in O-RAN can be categorized as shown in Fig. \ref{fig:fig1}. Vertical conflicts occur between applications at different layers, such as an xApp contradicting the policies enforced by an rApp. In contrast, horizontal conflicts emerge between xApps operating at the same layer, often competing for shared resources. Intra-component conflicts occur between xApps at the same Near-RT RIC, while inter-component conflicts concern xApps at different Near-RT RICs that control neighboring RAN nodes. Intra-component conflicts are further classified as direct, indirect, or implicit. A direct conflict arises when two xApps take different actions on the same parameter; for instance, one xApp increases transmission power to improve throughput, while another reduces it to save energy. Indirect conflicts occur when one xApp’s actions inadvertently affect another's objectives; for example, an xApp optimizing handovers might unintentionally reduce available resources for a neighboring cell managed by another xApp. Implicit conflicts between xApps' decisions are harder to detect because, while individually aligned with their specific objectives, they result in an undesirable overall network state. For instance, an xApp prioritizing latency implicitly violates another xApp aimed at load balancing. Implicit conflicts usually cannot be detected in advance; rather, they become apparent during long-term network operation.  

\subsection{Related Work}\label{subsec:related}

Conflict detection and resolution have gathered significant attention in the context of O-RAN, with many studies addressing the issue from architectural and algorithmic perspectives. A recent study proposed an architectural conflict mitigation framework embedded within the Near-RT RIC, which identifies and resolves conflicts using predefined rules and real-time monitoring, aligning with O-RAN's modular architecture principles \cite{adamczyk2023conflict}. Another study introduced a pre-emptive conflict detection architecture within the Service Management and Orchestration (SMO) framework, focusing on early identification of potential conflicts before they escalate to performance degradation. Their solution emphasizes proactive monitoring of multi-xApp interactions \cite{armstrong2024pre}.

Algorithmic approaches have also been extensively studied for conflict mitigation in O-RAN. The authors in \cite{wadud2024xapp} proposed a mobility-driven energy-saving approach to manage xApp-level conflicts, particularly those arising from overlapping objectives, such as energy efficiency and mobility management. Another study in \cite{wadud2023conflict} employed a game-theoretic framework within the Near-RT RIC to model and resolve conflicts among xApps competing for shared network resources. Additionally, another study presented a team-based learning framework to promote collaborative decision-making among xApps, reducing implicit and indirect conflicts during resource allocation \cite{zhang2022team}. This work indirectly addresses conflicts by promoting cooperative learning among xApps for efficient resource distribution. However, collaboration among xApps presents high complexity in real scenarios where multiple vendor-specific xApps are considered. While these studies provide valuable insights into conflict detection and resolution mechanisms to achieve conflict-free O-RAN operation, there is a notable lack of specific use case-based validations or proof-of-concept scenarios that quantitatively compare different conflict resolution policies under realistic network conditions.

\begin{table}
\centering
\caption{Acronyms}
\label{table1}
\setlength{\tabcolsep}{3pt}
\begin{tabular}{|p{45pt}||p{170pt}|}
\hline
\textbf{Acronym} & \textbf{Meaning} \\
\hline
API & Application Programming Interface\\
CD & Conflict Detector\\
CMF & Conflict Mitigation Framework\\
COMIX &  Conflict Management for Multi-Channel Power \par Control in O-RAN xApps\\
CP & Control Parameter\\
CQI & Channel Quality Indicator\\
CU & Central Unit\\
DC & Direct Conflict\\
DMPC & Downlink Multi-channel Power Control\\
DRM & Data Rate Maximization\\
DRL & Deep Reinforcement Learning\\
DU & Distributed Unit\\
E2SM & E2 Service Model\\
EE & Energy Efficiency\\
EES & Energy Efficiency-based Selection\\
EEVS & EE Violation-based Selection\\
FH & Front-haul\\
IC & Indirect Conflict\\
KPI & Key Performance Indicator\\
MaxTS & Maximum Throughput-based Selection\\
MinPS & Minimum Power-based Selection\\
MNO & Mobile Network Operator\\
NDT & Network Digital Twin\\
Near-RT RIC & Near-Real-Time RAN Intelligent Controller\\
Non-RT RIC & Non-Real-Time RAN Intelligent Controller\\
O-RAN & Open Radio Access Network\\
QoS & Quality of Service\\
RB & Resource Block\\
RIC & RAN Intelligent Controller\\
RU & Radio Unit\\
SLA & Service-level Agreement\\
SINR & Signal-to-Interference-plus-Noise Ratio\\
SMO & Service, Management and Orchestration\\
TVS & Throughput Violation-based Selection\\
UE & User Equipment\\
xApp & eXtended Application\\
\hline
\end{tabular}
\label{tab1}
\end{table}

\subsection{Paper summary and contributions}\label{subsec:contribution}

This work presents a conflict resolution management scheme for
multi-channel power control in O-RAN xApps (COMIX), designed to address conflicting objectives among RU-targeted decisions. Aligned with the standardized CMF defined by the O-RAN Alliance, we adopt a generalizable approach to classify and manage conflicts. Our focus is on resolving direct conflicts between two Deep Reinforcement Learning (DRL)-based xApps for multi-channel power control at the RU level in practical O-RAN scenarios. The main contributions of this work may be summarized as follows:

\begin{itemize}
    \item We formulate the power control problem for RUs in multi-channel, multi-cell, and mobility-aware multi-user O-RAN networks as a non-convex optimization problem. This formulation considers the impact of inter-channel and inter-cell interference on both experienced users' throughput and system energy efficiency (EE). To address this, we propose DRL algorithms to solve two variants of the problem, adhering to RU power budget and resource block (RB) constraints while ensuring Quality of Service (QoS) requirements.
    \item Two DRL-based xApps are developed for intelligent power regulation at the RU level. Leveraging E2 O-RAN metrics, the first xApp aims to maximize total system throughput, while the second focuses on optimizing overall EE. Both xApps dynamically adjust RU power levels under multi-channel transmissions, highlighting the inherent conflict in their objectives. We adopt and compare multiple decision-making policies applied by the CMF to resolve antagonistic conflicts by jointly considering different policies such as Service-Level Agreement (SLA) violations, power budget constraints, and QoS satisfaction.
    \item In the COMIX framework, we provide a structured methodology for handling competitive suggestions from xApps with conflicting objectives. To improve the precision of the decision, the proposed framework leverages the Network Digital Twin (NDT) \cite{karamplias2022towards, skianis2023digital} to evaluate the impact of candidate actions before final decision, ensuring minimal disruption of live network operations. An analytical and general-purpose workflow is demonstrated to guide the application of COMIX in realistic O-RAN scenarios.
\end{itemize}

For ease of exposure, Table \ref{table1} summarizes all the acronyms used throughout this article.


\section{Architectural System Model}\label{sec:systemmodel}

\begin{figure}[t]
\centering
\includegraphics[trim={0.5cm 0.4cm 0.5cm 0.5cm},clip,width=1\columnwidth]{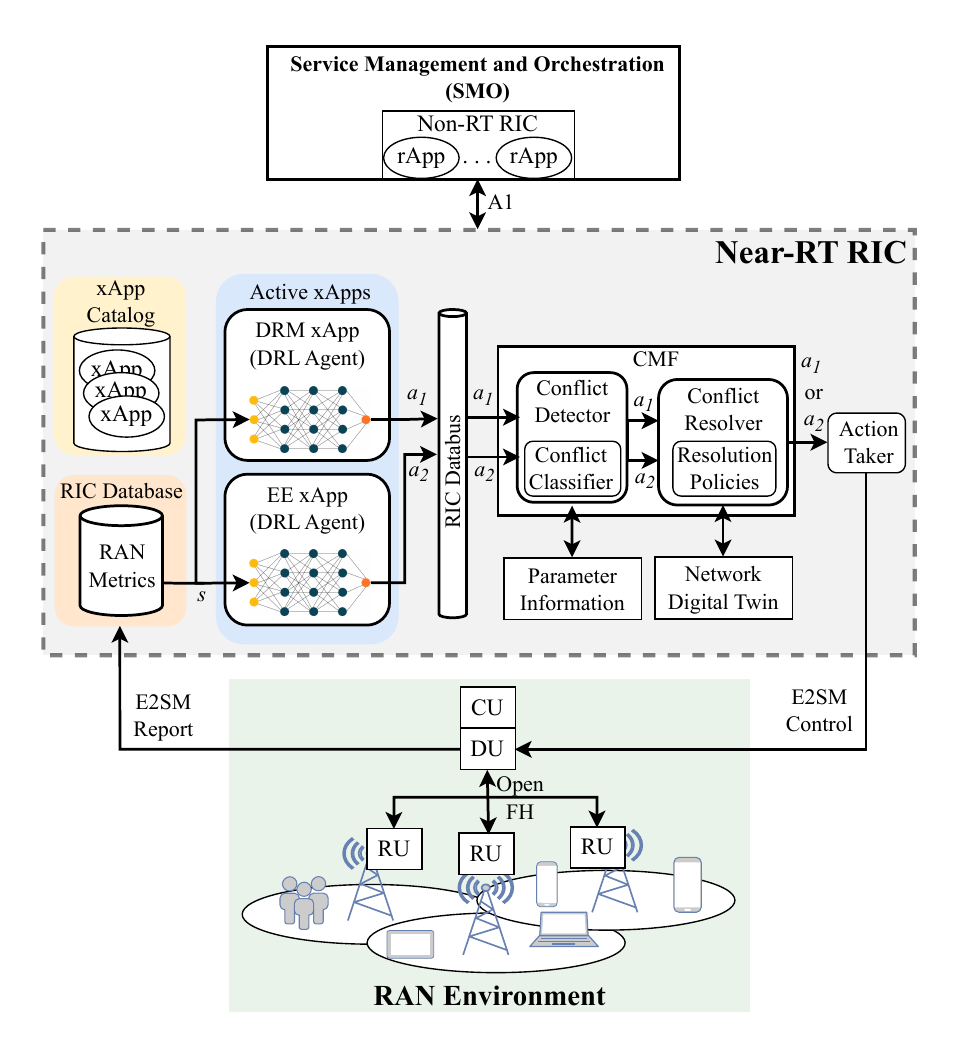}
\caption{General O-RAN architecture considered for COMIX.}
\label{fig:fig2}
\end{figure}

This section outlines the overall architecture for realizing an end-to-end and O-RAN-compliant optimization scenario, including O-RAN metrics collection, multi-xApp models' inference, CMF-driven conflict detection and resolution, and final action taking \cite{garcia2021ran}. The architecture considered in COMIX is shown in Fig. \ref{fig:fig2}. The key components of the architecture and their role are described below.

\subsection{Components and their Roles}

Firstly, the \textit{SMO} hosts the \textit{Non-RT RIC}, which operates at long timescales to provide high-level guidance and configuration to the \textit{Near-RT RIC} through the A1 interface. This includes hosting rApps for policy generation, performance monitoring, and non-real-time optimization. The A1 interface enables the exchange of policy and configuration data between the Non-RT RIC and Near-RT RIC, ensuring alignment between strategic and real-time control actions. For instance, vendor-specific or network-defined policies may be enforced from the Non-RT RIC to the Near-RT RIC through A1, imposing xApps and/or CMF component to comply with these policies (e.g. the system-level energy consumption must be below a certain threshold, or the experienced data rate by certain users should exceed a threshold) \cite{dai2024ran}. Such policies are integrated within Conflict Resolver (noted as \textit{Resolution Policies}) to be taken into account when potential conflicts are evaluated.

The Near-RT RIC lies at the core of the architecture and enables real-time optimization and conflict management through its key components:
\begin{enumerate}
    \item \textbf{xApp Catalog:} Contains a repository of all available xApps originated by different vendors. xApps can be activated or instantiated based on network requirements. For example, a Data Rate Maximization (DRM) xApp and an EE xApp may be active at a certain period for purposes of maximizing the users' throughput and system EE, respectively \cite{dinh2025demystifying}.
    \item \textbf{Active xApps:} All xApps that are active to provide control decisions (e.g. DRM and EE xApps). These xApps may leverage supervised, unsupervised, or DRL algorithms to optimize RAN resources (e.g., regulate RU power levels for DRM). In the COMIX scenario, the DRM xApp focuses on maximizing system throughput, while the EE xApp optimizes energy efficiency. Both xApps receive an O-RAN state vector $s$ and provide action suggestions ($a_1$ and $a_2$) or control decisions, which are further evaluated by the \textit{CMF} component for potential conflicts. Note that the CMF may be fed with multiple candidate actions when more xApps are active in order to assess conflicting xApp groups. 
    \item \textbf{RIC Database:} Maintains RAN metrics and potential historical performance data gathered from the network through the E2 Service Model (E2SM) interface. These data are accessible as E2SM Reports to both active xApps and CMF for informed decision making.
    \item \textbf{RIC Databus:} The messaging layer for data provisioning within Near-RT RIC. Usually, CMF is subscribed to RIC Databus for purposes of receiving or consuming new data from the xApps output. xApps often act as data providers for the RIC Databus, bridging communication with the CMF.
    \item \textbf{CMF Component:} The CMF is responsible for detecting, classifying, and resolving conflicts between xApps. It consists of: 
    \begin{itemize}
        \item \textit{Conflict Detector} which identifies and classifies conflicts (e.g. direct, indirect, or implicit) between the xApps' proposed actions based on predefined rules and historical data. This component uses \textit{Parameter Information} data to define conflicts such as the recently changed parameters and the parameters groups (a parameters group is defined as a cluster of configuration parameters that affect the same KPI). The internal functionality of the Conflict Detector, called \textit{Conflict Classifier}, is further described in Section \ref{generalCD}.
        \item \textit{Conflict Resolver} which employs various Resolution Policies, such as priority-based, SLA-based, data rate-targeted, or fairness-driven approaches, to resolve conflicts. The resolution process also integrates the NDT, which simulates the impact of candidate actions on network performance before selecting the optimal action to execute \cite{salehi2024multiverse}. NDT-driven conflict resolution is advantageous against any other reactive mechanism, which evaluates the decisions after their application on the live network, allowing pro-active knowledge about each decision's impact.
    \end{itemize}
    \item \textbf{NDT:} The NDT is a virtualized model of the live network that predicts the effects of xApp actions by applying the all candidate decisions on the simulated environment. In the case of DRL models, the NDT can simply use the same environment as used during the training to evaluate different candidate corrective actions. By evaluating potential actions through KPIs in a simulated environment (e.g., overall EE and users' data rate derived upon $a_1$ and $a_2$ actions), the NDT improves decision accuracy pre-action, minimizing the risk of adverse impacts on the live network. Note that the absence of NDT capabilities forces the Conflict Resolver to implement non-guided actions (based on predefined rules or patterns), potentially leading to network reconfiguration when the action is degradative, further imposing network disruption.
    \item \textbf{RAN Environment:} Consists of the E2 RAN nodes, namely the RUs, DUs and CUs. RUs are managed through the Open Front Haul (Open FH) interface, enabling flexible, multi-vendor configurations. The RUs serve UEs and are directly influenced by the actions taken by the Near-RT RIC. Note that the UE traffic pattern, UE metrics, UE demands, UE mobility, UE handovers, inter-cell interferences, channel losses, and antenna system parameters (transmitting power levels, UE/RU association, physical resource blocks) are considered part of the RAN environment. Actions imposed by the Near-RT RIC's \textit{Action Taker} are conveyed over E2 interface as E2SM Control messages, whereas the RAN metrics acknowledged to the RIC Database are collected as E2SM Report messages over E2.
\end{enumerate}

\subsection{Generalized xApp Conflict Classification}\label{generalCD}

\begin{figure}[t]
\centering
\includegraphics[trim={0.8cm 0.4cm 0.5cm 0.4cm},clip,width=\columnwidth]{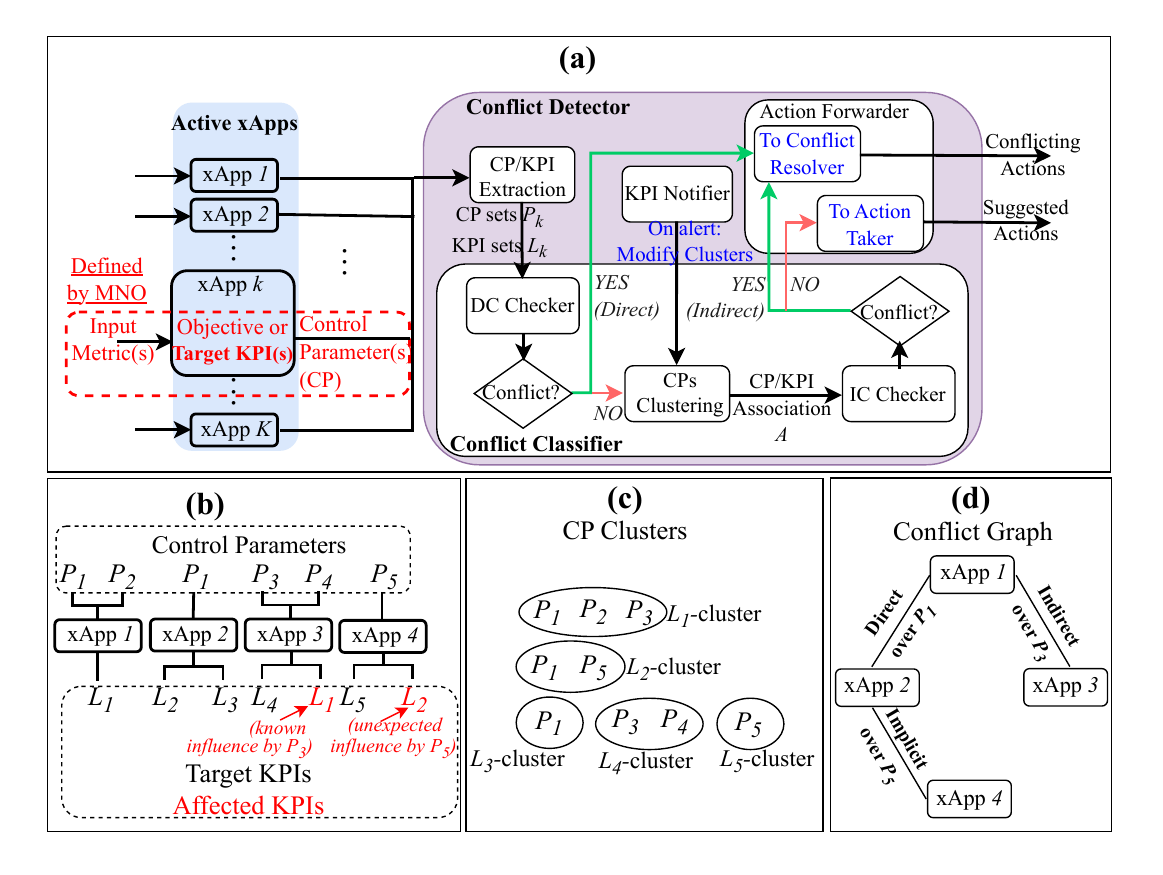}
\caption{Generalized Direct/Indirect Conflict Detector. \textbf{(a)} Internal Architecture; \textbf{(b)} Example of CPs/KPIs associated with 4 xApps; \textbf{(c)} CP Clusters per KPI; \textbf{(d)} Conflict Graph.}
\label{fig:fig3}
\end{figure}

This section outlines a general-purpose \textit{Conflict Detector} (CD) used in the COMIX framework. For the ease of exposure, Table \ref{table2} lists all the mathematical notations used in this work. The role of the CD is to identify pro-actively (i.e. beforehand) direct conflicts (DCs) and indirect conflicts (ICs). Since implicit conflicts can only be detected upon unexpected KPI degradation occurrences, CD supports the identification of implicit conflicts re-actively (i.e. after their negative impact). As shown in Fig. \ref{fig:fig3}a, CD contains several internal logical modules to analyze, classify and forward potential contradicting decisions made by different xApps. Let $\mathcal{K} = \{1,2,\dots,K\}$ denote the set of active xApps identifiers at a given time slot. Each xApp $k \in \mathcal{K}$ is characterized by its input metrics which are used for inferring the xApp model to obtain output Control Parameters (CPs), referred to as xApp decisions. Decisions about CPs configuration targets at optimizing a specific objective or a target KPI (e.g. power consumption below a threshold), according to which the xApp model has been trained. In some cases, an xApp might provide CPs that affect not only its target KPIs, but also KPIs of other xApps (e.g. EE-targeted power control of RUs by one xApp affects the data rate of UEs, which might be the objective of another xApp). In this sense, we assume that Mobile Network Operator (MNO) -which serves as the xApp owner- is responsible for predefining the xApp inputs, CPs, and all \textit{a priori} known KPIs that are affected by the CPs \cite{zolghadr2024learning}. Once the active xApps complete their model inference, all the CP decisions enter the CD (via the RIC Databus) for potential identification of contradicting proposed actions. The CD internal architecture is structured as follows:

\begin{enumerate}
    \item \textbf{CP/KPI Extraction:} This module extracts the basic information (i.e. CPs and KPIs) from each active xApp $k \in \mathcal{K}$. Firstly, it identifies the different parameters and KPIs received from all xApps, denoting $\mathcal{P}=\{P_1,P_2,\dots,P_X\}$ as the CP set and $\mathcal{L}=\{L_1,L_2,\dots,L_Y\}$ as the KPI set. As output, it simply provides the CP set $\mathcal{P}_k \subseteq \mathcal{P}$ (i.e. CPs controlled by xApp $k$) and the KPI set $\mathcal{L}_k \subseteq \mathcal{L}$ (i.e. KPIs affected by xApp $k$).
    \item \textbf{DC Checker:} Occurrences of direct conflicts between xApps are checked by the DC Checker. The tests involved in this module search for combinations of xApps with at least one common parameter. Thus, the DC Checker loops over all CPs and returns the xApp groups with common CPs, indicating direct conflicts over specific control parameters. The xApp decisions with direct conflicts are passed to the \textit{Action Forwarder}, whereas CP/KPI sets are also forwarded to the \textit{CPs Clustering} module for further indirect conflict investigation.
    \item \textbf{CPs Clustering:} This module creates a cluster for each KPI $L_y \in \mathcal{L}$ which contains all the CPs that affect $L_y$ ($y=1,2,\dots,Y$). Clusters are formed based on the Association Matrix $A$ (dimensions $X \times Y$) with elements $A_{i,j} = 1$, when CP $P_i$ affects KPI $L_j$, or $A_{i,j} = 0$, otherwise.
    \item \textbf{IC Checker:} Indirect conflicts are detected by the IC Checker, which searches for xApp groups with CPs belonging in the same cluster. In this way, IC Checker is able to detect which xApps influence common KPIs and, in case of IC, xApps decisions will be sent to the Conflict Resolver, otherwise they will be directly passed to Action Taker due to conflict absence. 
    \item \textbf{Action Forwarder:} This module establishes the proper end-points or interfaces (e.g. by exposing APIs) to forward the xApp decisions towards either the Conflict Resolver (for contradicting decisions) or the Action Taker (for conflict-free decisions). Hence, if DCs or ICs are raised by the DC or IC Checkers, xApp decisions are passed to the component called \textit{To Conflict Resolver}, otherwise they are passed to the component called \textit{To Action Taker}.
    \item \textbf{KPI Notifier:} To partially support the detection of implicit conflicts, the proposed CD includes a KPI Notifier. This module keeps track of KPI monitoring data (e.g. stored in RIC Database) and raises alarms when an unexpected KPI degradation is observed due to the control of a specific CP. Thus, if an alert is triggered, KPI Notifier enforces a manual modification of the CP clusters by extending the CP group affecting the degraded KPI.
\end{enumerate}

\begin{table}
\centering
\caption{Mathematical symbols}
\label{table2}
\setlength{\tabcolsep}{3pt}
\begin{tabular}{|p{30pt}p{83pt}||p{24pt}p{85pt}|}
\hline
\textbf{Symbol} & \textbf{Meaning} & \textbf{Symbol} & \textbf{Meaning} \\
\hline
$\mathcal{K}$ & Set of active xApps & $K$ & Number of active xApps\\
$\mathcal{P}$ & Set of all CPs & $P_i$ & Control parameter $i$\\
$\mathcal{L}$ & Set of all KPIs & $L_j$ & KPI $j$\\
$X$ & Total number of CPs & $Y$ & Total number of KPIs\\
$\mathcal{P}_k$ & Set of CPs \par controlled by xApp $k$ & $\mathcal{L}_k$ & Set of KPIs \par affected by xApp $k$\\
$A$ & Association matrix & $A_{i,j}$ & Impact index ($1$ if CP $P_i$ \par affects KPI $L_j$, else $0$)\\
$\mathcal{N}$ & Set of RUs & $N$ & Number of RUs\\
$\mathcal{M}$ & Set of RBs & $M$ & Number of RBs\\
$W_m$ & Bandwidth of RB $m$ & $s$ & Service identifier\\
$\mathcal{U}$ & Set of UEs & $U$ & Total number of UEs\\
$D$ & Demand vector of all UEs & $d_u$ & Demand of UE $u$\\
$A_{n,m,u}$ & Association index ($1$ if \par UE $u$ is served by \par RB $m$ of RU $n$, else $0$) & $p_{n,m}$ & Power level allocated \par to RB $m$ of RU $n$\\
$\Bar{P}$ & Maximum power RUs & $\Bar{p}$ & Minimum power of RBs\\
$\gamma_{n,m,u}$ & SINR of UE $u$ over \par RB $m$ of RU $n$ & $n_0$ & Received noise power \par at the UE\\
$L_{n,m,u}$ & Path loss between UE $u$ \par and RU $n$ over RB $m$ & $F_1$ & System data rate\\
$r_{n,m,u}$ & Data rate of UE $u$ over \par RB $m$ of RU $n$ & $F_2$ & System EE\\
$\mathcal{T}$ & Set of time slots & $T$ & Number of time slots\\
$\boldsymbol{p}$ & Power vector of all \par RUs and RBs & $\xi$ & Throughput coefficient \par in EE xApp\\
$CQI_{n,m}$ & CQI of RB $m$ of RU $n$ & $\boldsymbol{s}(t)$ & State vector at time $t$\\
$\boldsymbol{V}(t)$ & Variation vector at time $t$ & $P_s$ & Power step\\
$\boldsymbol{r_1}(t)$ & Reward received at \par time $t$ by DRM xApp & $\boldsymbol{a_1}(t)$ & Action proposed at \par time $t$ by DRM xApp\\
$\boldsymbol{r_2}(t)$ & Reward received at \par time $t$ by EE xApp & $\boldsymbol{a_2}(t)$ & Action proposed at \par time $t$ by EE xApp\\
\hline
\end{tabular}
\label{tab2}
\end{table}

To concretely explain the CD information flow and components' functionality, let us consider the example of Fig. \ref{fig:fig3}b, where 4 active xApps are considered.  The set of all CPs is $\mathcal{P} = \{P_1,P_2,P_3,P_4,P_5\}$ and the set of all KPIs is $\mathcal{L} = \{L_1,L_2,L_3,L_4,L_5\}$. Initially, the CP/KPI Extraction component defines the CP/KPI sets per xApp as: $\mathcal{P}_1 = \{P_1,P_2\}$, $\mathcal{L}_1 = \{L_1\}$, $\mathcal{P}_2 = \{P_1\}$, $\mathcal{L}_2 = \{L_2,L_3\}$, $\mathcal{P}_3 = \{P_3,P_4\}$, $\mathcal{L}_3 = \{L_4,L_1\}$, $\mathcal{P}_4 = \{P_5\}$, $\mathcal{L}_4 = \{L_5\}$. For the ease of exposure, we assume that, initially, it is not known that $P_5$ affects KPI $L_2$. Noteworthy, KPI $L_1$ is a target KPI of xApp 1, but is unavoidably affected by the operation of xApp 3 due to the variation of $P_3$. Then, the DC Checker tests for common elements (i.e. same CPs regulated by different xApps) by computing the intersections across the CP sets. In this example, there is a DC between xApp 1 and xApp 2 since $\mathcal{P}_1 \cap \mathcal{P}_2 = \{P_1\}$, meaning that both control $P_1$. Upon DC detection, the decisions of xApp 1 and xApp 2 are passed to the Conflict Resolver, whereas all CP/KPI sets are further forwarded to the CPs Clustering. The latter forms the following clusters: $L_1\text{-cluster} = \{P_1,P_2,P_3\}$, $L_2\text{-cluster} = \{P_1\}$, $L_3\text{-cluster} = \{P_1\}$, $L_4\text{-cluster} = \{P_3,P_4\}$, $L_5\text{-cluster} = \{P_5\}$. Based on these clusters (see Fig. \ref{fig:fig3}c), IC Checker identifies that $P_3$ is a member of $L_1$\text{-cluster}, so it affects the operation of xApp 1 while being only a direct CP of xApp 3. Thus, the decisions of xApp 1 and xApp 3 form an IC and are passed to the Conflict Resolver, whereas xApp 4 decision is passed directly to the Action Taker. At a later time slot, KPI Notifier can trigger an alarm due to an unexpected degradation of KPI $L_2$ attributed to $P_5$. This in turn enforces a modification of $L_2$-cluster by adding $P_5$. Including $P_5$ in $L_2$-cluster leads the IC Checker to detect an IC between xApp 2 and xApp 4. Thus, KPI Notifier allows for detecting implicit conflicts over the long-term network operation, which are finally identified as ICs. Fig. \ref{fig:fig3}d depicts the overall conflict graph, including the implicit conflict detected later in time (relative to the proactively detected DC/IC conflicts), after observing an unexpected degradation of $L_2$.

\section{Design of xApps}

This section describes the mathematical background for designing and training two DRL-based algorithms used as Near-RT RIC xApps. Both xApps operate in the same multi-cell RAN environment, aiming to provide multi-channel power control policies for O-RAN RUs.

\subsection{Downlink Multi-channel Power Control}\label{subsec:powercontrol}

Solving the optimization problem of Downlink Multi-channel Power Control (DMPC) involves finding the proper power-level configuration of all RUs so as to optimize a specific objective, under power constraints, user demands, mobile traffic, inter-channel and inter-cell interferences. We consider an O-RAN wireless network consisting of $N$ RUs from the set $\mathcal{N}=\{1,2,\dots,N\}$, where each RU serves a coverage area. In this setup, the operational frequency band is segmented into multiple RBs based on a given channel segmentation. Each RU supports $M$ available RBs (from the set $\mathcal{M}=\{1,2,\dots,M\}$) for downlink physical-layer transmissions, where each RB $m \in \mathcal{M}$ has a specific bandwidth $W_m$. A complete frequency reuse scheme is assumed across the O-RAN transmissions, meaning all RUs operate on the same frequency band, introducing inter-cell interferences over common-RB transmissions.

Each UE in the network, denoted as $u \in \mathcal{U}$ (where $\mathcal{U}=\{1,2,\dots,U\}$), requests a specific service $s$ based on its SLA profile in order to meet its QoS requirements. Each service $s$ corresponds to a particular throughput demand, expressed in Mbps. To represent these demands, a demand vector $D$ with elements $d_u$ is used to notify the requested service class of each user $u$. In the proposed system model, each UE $u$ is associated with a single RB $m$ from a specific RU $n$, determined by the binary association matrix $A$, where each element $A_{n,m,u} = 1$ if user $u$ is associated with RB $m$ of RU $n$, or $0$ otherwise. The power level allocated to RB $m$ of RU $n$ is represented as $p_{n,m}$. To ensure power efficiency, the system imposes a sum-power constraint on each RU $n$, as given by $\sum_{m=1}^M p_{n,m} \leq \Bar{P}$, where $\Bar{P}$ is the maximum total transmitting power allowed for each $RU$. Additionally, a minimum power level constraint $p_{n,m} \geq \Bar{p}$ is enforced to account for signaling and sleeping operations of each RB. Also, each RU can cover multiple users.

The Signal-to-Interference-plus-Noise Ratio (SINR) experienced by a user $u$, associated with RB $m$ of RU $n$, is given by:

\begin{equation}
    \gamma_{n,m,u} = \frac{p_{n,m} L_{n,m,u}}{\sum_{t \neq k} p_{t,m} L_{t,m,u} + n_0}
\end{equation}

\noindent where $L_{n,m,u}$ denotes the channel losses between RU $n$ and UE $u$ over RB $m$, and $n_0$ represents the received noise power at the UE. The achievable data rate 
$r_{n,m,u}$ for a user $u \in \mathcal{U}$ associated with RB $m$ of RU $n$ is computed using the Shannon formula:

\begin{equation}
    r_{n,m,u} = W_m \cdot \log_2(1+\gamma_{n,m,u})
\end{equation}

The user association is determined so as to ensure that each UE receives the highest achievable rate. Note that the system-level data rate is defined as:

\begin{equation}\label{obj1}
    F_1 = \sum_{n=1}^N\sum_{m=1}^M\sum_{u=1}^U A_{n,m,u} \cdot r_{n,m,u}
\end{equation}

Similarly, the system-level EE is defined as:

\begin{equation}\label{obj2}
    F_2 = \frac{\sum_{n=1}^N\sum_{m=1}^M\sum_{u=1}^U A_{n,m,u} \cdot r_{n,m,u}}{\sum_{n=1}^N\sum_{m=1}^M p_{n,m}}
\end{equation}

\subsection{Optimization Problem via DRL}\label{subsec:opt}

\begin{figure}[t]
\centering
\includegraphics[trim={0.5cm 0.4cm 0.5cm 0.5cm},clip,width=0.95\columnwidth]{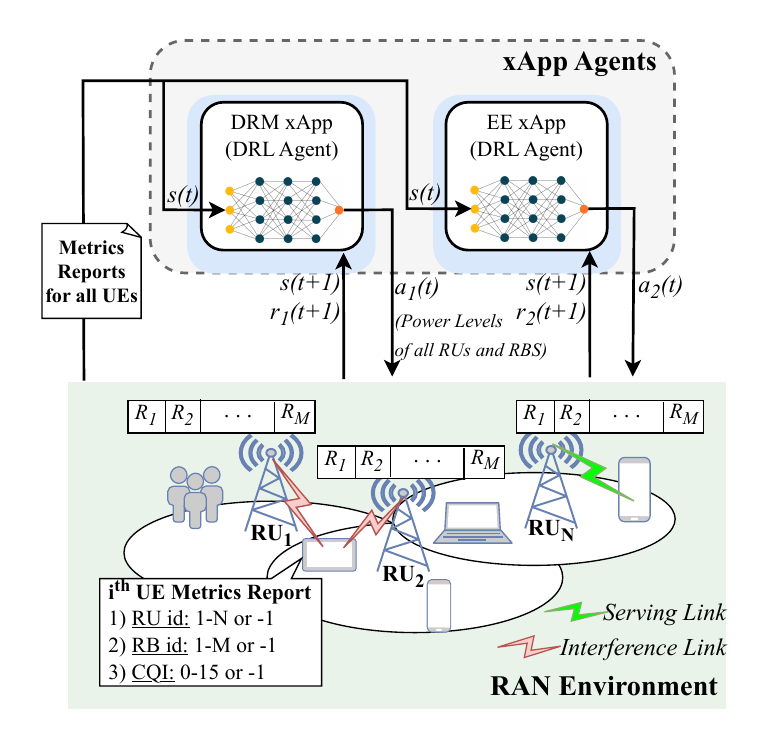}
\caption{DRL cycles between xApps and O-RAN environment.}
\label{fig:fig4}
\end{figure}

To regulate power control across RUs intelligently, the DMPC problem is solved using a DRL framework, implemented as two distinct xApps with different objectives: (i) the DRM xApp aims to optimize power allocation to maximize the total system throughput for UEs \cite{giannopoulos2023learning}; (ii) the EE xApp selects the power allocation to minimize energy consumption while maintaining QoS requirements \cite{spantideas2021joint}.

The intelligent agent running on the Near-RT RIC adheres to the following constraints: (i) respecting the total power budget $\Bar{P}$ of RUs and the minimum power limitations $\Bar{p}$ of RBs; (ii) considering the available RBs for each RU; (iii) satisfying the QoS demands specified by each UE's SLA profile. Each UE reports the following metrics to the Near-RT RIC's Database: Associated RU ID, Associated RB ID, Current data rate in Mbps. Fig. \ref{fig:fig4} shows the DRL interaction cycles between the DRM/EE xApps and the RAN environment.

The simultaneous operation of DRM and EE xApps introduce conflicts, so the goal of the COMIX scheme is to ensure that RUs dynamically adjust their power levels while balancing the conflicting objectives based on predefined resolution policies. These policies are enforced by the Non-RT RIC and are applied by the Conflict Resolver. Below, we formulate the optimization problem defined for each xApp, and we determine the DRL elements (state, action, reward) considered in the solution \cite{giannopoulos2021deep}.

\subsubsection{Data Rate Maximization xApp}\label{subsubsec:through}

The DRL-based DRM xApp focuses on solving the following optimization problem:

\begin{equation}\label{problem1}
\argmax_{\boldsymbol{p}} \Bigl( \frac{1}{T}\sum_{t=1}^T F_1(\boldsymbol{p}) \Bigr)
\end{equation}

\begin{align*}
\text{s.t.} \quad & \text{(C1)} \quad \sum_{m=1}^M p_{n,m} \leq \Bar{P}, \quad \forall n \in \mathcal{N} \\
& \text{(C2)} \quad p_{n,m} \geq \Bar{p}, \quad \forall n \in \mathcal{N}, m \in \mathcal{M} \\
& \text{(C3)} \quad \sum_{n=1}^N \sum_{m=1}^M A_{n,m,u} \leq 1, \quad \forall u \in \mathcal{U} \\
& \text{(C4)} \quad r_{n,m,u} \geq d_u, \quad \forall u \in \mathcal{U}
\end{align*}

\noindent where it is implied that DRM xApp looks for a power vector $\boldsymbol{p}=\{p_{n,m}\}$ (with dimensions $N \times M$) such that to maximize the expected (long-term) system-level data rate after a series of $T$ training episodes \cite{ahmadian2024long}. 

The following DRL elements are considered by the DRM xApp during the training process:

\textbf{State:} Considering the set of time slots $\mathcal{T}=\{1,2,\dots,T\}$, the DRL agent observes a state vector $\boldsymbol{s}(t)$ at each time $t \in \mathcal{T}$. The state vector is defined as $\boldsymbol{s}(t) = \{CQI_{n,m}\}$ (dimensions $N \times M$), where each element $CQI_{n,m}$ denotes the channel quality indicator (CQI) of the UE that is served by RB $m$ of RU $n$. CQI values vary between $0-15$ and represent the quantized version of the SINR values $\gamma_{n,m,u}$ into 16 discrete levels (i.e. zero CQI corresponds to poor SINR levels, CQI at 15 means perfect propagation conditions), as defined in \cite{giannopoulos2021deep}. When no UE is associated with a certain RB $m$ of RU $n$, we note $CQI_{n,m} = -1$ by convention.

\textbf{Action:} After observing $\boldsymbol{s}(t)$, DRL agent provides a control decision on the RUs power configuration by assigning a new power vector $\boldsymbol{a_1}(t) = \boldsymbol{p}(t) = \boldsymbol{p}(t-1) + \boldsymbol{V}(t)$. Specifically, the new power vector is equal to previous-slot power vector changed by the variation vector $\boldsymbol{V}(t)$. The latter is formed by selecting one RB per RU and either increase or decrease its transmitting power by a predefined power step $P_{s}$. For instance, the variation vector:

\begin{equation}
    \boldsymbol{V}(t) = \begin{bmatrix}
        0 & +P_s & \cdots & 0 \\
        -P_s & 0 & \cdots & 0 \\
        \vdots & \vdots & \ddots & \vdots \\
        0 & 0 & \cdots & +P_s
        \end{bmatrix}_{N \times M}
\end{equation}

\noindent reflects that RB $2$ of RU $1$ and RB $M$ of RU $N$ will be increased by $P_s$ (in Watts), whereas the RB $1$ of RU $2$ will be decreased by $P_s$. The power level of the rest RBs will be the same as in the previous-slot power level. In this formulation, DRL agent is able to apply smooth changes in the power configuration across RUs to optimize objective $F_1$.  

\textbf{Reward:} After applying an action $\boldsymbol{a_1}(t)$, the DRL agent receives a scalar reward $\boldsymbol{r_1}(t+1)$ reflecting the extent to which the action is beneficial relative to the objective. In the case of DRM xApp, the reward is defined as $\boldsymbol{r_1}(t+1) = F_1(t) - F_1(t-1)$, where $F_1(t)$ and $F_1(t-1)$ are derived by applying \eqref{obj1} considering the power configuration $\boldsymbol{p}(t)$ and $\boldsymbol{p}(t-1)$, respectively. In this way, the DRL agent's reward reflects the data rate increment/decrement derived after applying $\boldsymbol{a_1}(t)$ .

\subsubsection{Energy Efficiency Maximization xApp}\label{subsubsec:eecontrol}

Similarly, the EE xApp aims to solve the following optimization problem:

\begin{equation}\label{problem2}
\argmax_{\boldsymbol{p}} \Bigl( \frac{1}{T}\sum_{t=1}^T F_2(\boldsymbol{p}) \Bigr)
\end{equation}

\begin{align*}
\text{s.t.} \quad & \text{(C1)-(C3)} \\
& \text{(C4')} \quad r_{n,m,u} \geq \xi \cdot d_u, \quad \forall u \in \mathcal{U}
\end{align*}

\noindent where EE xApp searches a power vector $\boldsymbol{p}$ for all RUs/RBs to ensure a maximized data rate vs power consumption ratio \cite{ullah2024sum}. Note that constraint (C4') implies that the allocated throughput of each UE can be decreased for purposes of ensuring power savings, given that $\xi \in [0,1]$ denotes a constant value that allows for reduced throughput allocation at the UE (e.g., $r_{n,m,u} \geq 0.8 \cdot d_u$). In this manner, DRM xApp ensures maximized throughput and full QoS satisfaction (due to (C4)) while ignoring the power consumption, whereas EE xApp guarantees low power consumption while allowing for a slight degradation of the requested throughput (due to (C4')) by a factor $(1-\xi)\%$. Overall, both xApps optimize the O-RAN from different perspectives, highlighting the full QoS satisfaction vs EE trade-off.

To provide a suboptimal solution, the EE xApp adopts the same state/action space as in DRM, but the rewarding function is given by $\boldsymbol{r_2}(t+1) = F_2(t) - F_2(t-1)$, representing the EE change due to the application of action $\boldsymbol{a}_2(t)$. In summary, both xApps observes the same environment state $\boldsymbol{s}(t)$ and propose different actions (i.e. $\boldsymbol{a}_1(t)$ and $\boldsymbol{a}_2(t)$) such that to optimize $F_1$ and $F_2$, respectively.

\subsection{Network Digital Twin Functionality}

The NDT is an integral component of the COMIX framework, providing the dynamic simulation environment necessary for computing RAN/UE metrics and training the DRL-based xApps. The NDT models the O-RAN network environment, enabling the DRL agents to interact with a realistic, yet simulated, representation of the physical-layer RAN conditions \cite{salehi2024multiverse}. In the case of DRL algorithms, NDT should only coincide with the environment in which the agents perform actions, and is not necessarily a large-scale representation of the whole network. In this sense, NDT can be deployed as different instances depending on the DRL algorithm, with each instance abstracting only the environment required by a given DRL agent \cite{nguyen2024digital}. This ensures that the xApps can learn optimal policies (e.g., for multi-channel power control in RUs) before deployment in the live network. 

In the COMIX framewrok, the NDT allows for flexible configuration of the simulation parameters to accurately represent the O-RAN environment, and can be parameterized on-demand. The key configurable parameters include:
\begin{itemize}
    \item The number and topology of cells within the network area, along with the spatial location of the RUs.
    \item The channel and path loss models to estimate the received signal strength type-specific cells (e.g., macro, micro, or pico-cells; urban, suburban, or rural cells) \cite{jiang20213gpp, 3gpp2017study}.
    \item The maximum power budget per RU and the minimum power level of each RB.
    \item The operating 5G frequency band, numerology (spectrum channelization), and the total bandwidth allocated for downlink transmissions \cite{3gpp2017study}.
    \item The initial number of UEs in the network area, their maximum velocities under a mobility model (e.g. random walk), and their potential movement across cell boundaries.
    \item The probability of new UEs entering the network area from the borders during a given time slot. 
\end{itemize}

\subsubsection{Parameterizing NDT during xApp training}

The NDT initializes the simulation by randomly placing UEs within the network and assigning initial power levels to each RB, respecting the constraints on maximum and minimum power budgets. It computes channel gains based on the relative distance between UEs and RUs, following the 3GPP specifications for 5G cells \cite{3gpp2017study}. Using this information, the NDT calculates key metrics such as SINR and data rate for each UE across all RU and RB combinations. The core functionalities of the NDT include: (i) Interference calculation to evaluate the accumulated interference from non-serving RUs for each UE; (ii) Channel metrics production such as SINR and data rate for all potential RU/RB associations; (iii) Dynamic association to assign UEs to RBs that maximize their throughput; (iv) UE Mobility through a pre-defined model. This dynamic modeling supports time-varying network conditions, including potential handovers, ensuring that the DRL agents are trained in realistic and variable scenarios.

At each simulation time step, the NDT generates detailed UE metrics that can be used to form the DRL state vector, including the CQI and the IDs of the associated RU and RB. By simulating diverse network conditions during the DRL training, the NDT enables the xApps to refine their policies in a controlled environment, ensuring they are well-prepared for deployment in live O-RAN systems.

\subsubsection{Parameterizing NDT during xApp inference and conflict resolution}

When the xApp model is inferred with real UE metrics collected via E2, NDT can be configured instantly using the current UE metrics (positions, serving RUs and RBs) and the O-RAN parameters (RU power levels, topology). Also, when potential xApp conflicts are addressed by the Conflict Resolver, xApps decisions can be directly applied on the current NDT instance to gather the new O-RAN metrics. The Conflict Resolver then uses these metrics to evaluate which xApp complies with the current resolution policy (e.g. which xApp leads to maximum EE).

\section{Simulation Results}\label{sec:results}

This section presents the validation results of the COMIX scheme. First, we demonstrate the impact of the key learning hyperparameters to ensure optimal convergence of individual xApps. Then, we assess the network performance in terms of target KPIs (i.e. system-wide data rate and EE) under different conflict resolution policies using multiple simulation scenarios. Finally, considering the pre-trained DRM and EE xApps in inference mode, we analytically present the end-to-end COMIX workflow in a general-purpose manner, explaining the information flow step-by-step and showcasing the role of CMF and NDT components in addressing xApp-level conflicts. The constant hyperparameters used in the simulations are tabulated in Table \ref{tab:hyperparameters}. 

\begin{table}[t]
\centering
\caption{Simulated Environment and DRL Configuration}\label{tab:hyperparameters}
\begin{tabular}{|l||l|}
\hline
        \textbf{Parameter} & \textbf{Value} \\
        \hline
        Number Of RUs ($N$) & 3\\
        Number Of RBs ($M$) & 8\\
        Number Of UEs ($U$) & 12\\
        Type of cells & Urban\\
        Maximum RU Power ($\Bar{P}$) & 80 W\\
        Minimum RB Power ($\Bar{p}$) & 0.1 W\\
        UE demands ($d_u$) & $1-10$ Mbps \\
        Throughput coefficient in EE xApp ($\xi$) & $0.8$\\
        Training Episodes ($T$) & 2000\\
        Time Steps Per Episode & 200\\
        UE Velocity & 5 \text{m/s} \\
        UE Move Probability & 50\%\\
        Batch Size & 32 \\
        Replay Memory Capacity & 5000 \\
        Target $Q$-Network Update Frequency & 6 \\
        \hline
\end{tabular}
\end{table}

\subsection{Training Performance of xApps}\label{subsec:training}

Extensive simulations were carried out to ensure convergence of the algorithms. Both DRM and EE xApps were trained in identical environment setup after 2000 training episodes. The xApps only differed in the rewarding function, as specified in \eqref{obj1} and \eqref{obj2}, respectively. Although acting on the same environment with identical available actions, different rewards led the xApps to learn completely different policies. 

\begin{figure*}[t]
\centering
\includegraphics[width=\textwidth]{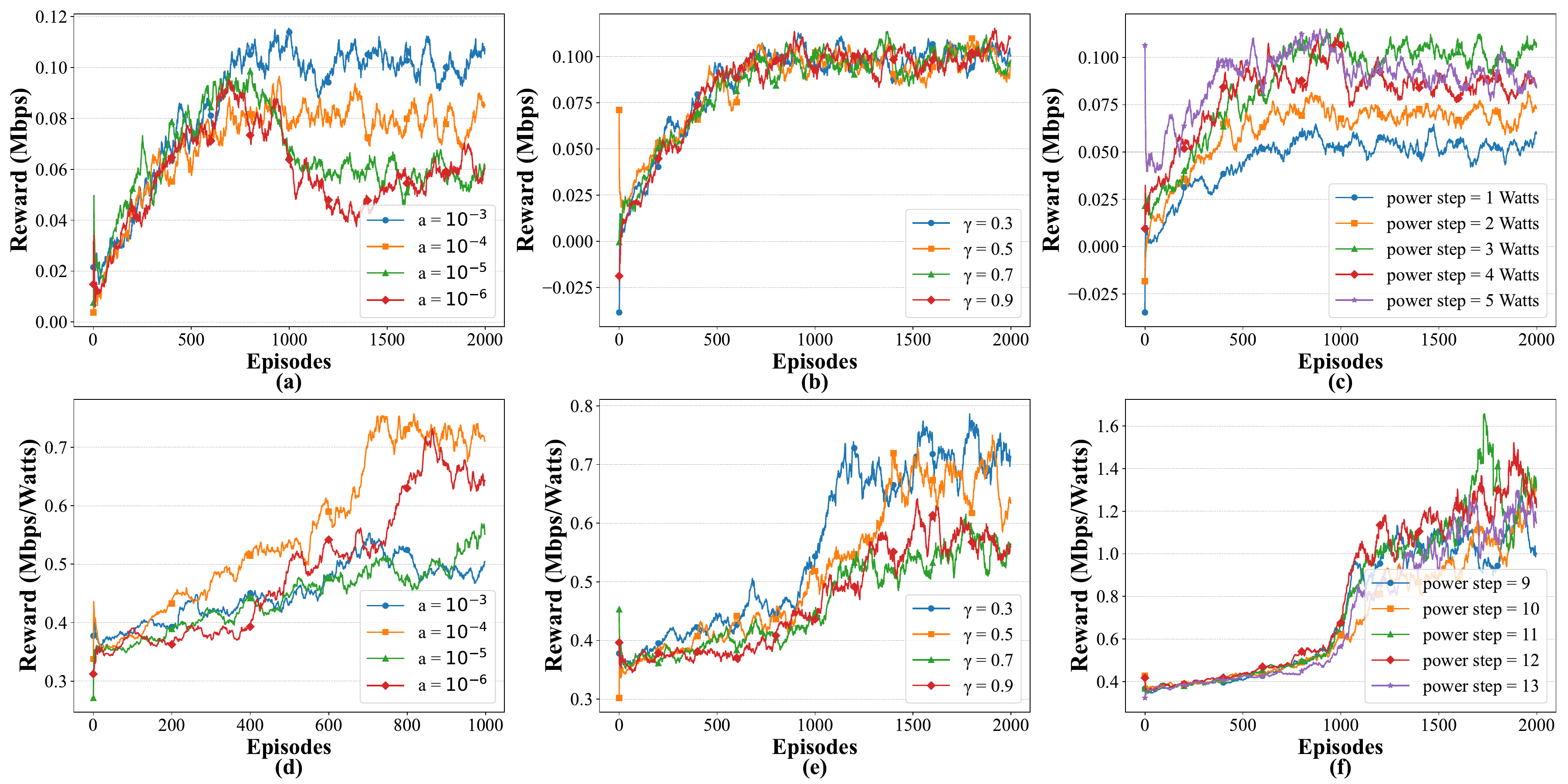}
    \caption{Learning curves of DRM \textbf{(a-c)} and EE \textbf{(d-f)} xApps for different learning rates (column 1), discount factors (column 2), and power steps (column 3).}
    \label{fig:learning}
\end{figure*}

To guarantee optimal convergence in DRL algorithms, the critical hyperparameters should be stabilized. Upon experimentation on different values of learning parameters, the most impactful hyperparameters were deemed to be the learning rate $\alpha$ (affects the intensity of weight update during backpropagation) and the discount factor  $\gamma$ (balances the trade-off between immediate and future rewards). Additionally, as a problem-specific hyperparameter, different values of the power step were tested, as they affect the granularity of RU power regulation. To assess the impact of each hyperparameter, we depict the learning curve of each DRL configuration, which reflects the reward time-course as a function of the training episodes. Fig. \ref{fig:learning}a-c shows the reward of different DRM xApp configurations in terms of the overall achieved data rate (in Mbps), whereas Fig. \ref{fig:learning}d-f depicts the same curves for EE xApp, representing the reward in terms of the system-wide EE (in Mbps/Watt).

Evidently, the discount factor exhibits no impact on the convergence of DRM xApp, whereas it strongly influences the convergence of EE xApp, with $\gamma=0.3$ presenting the best EE. The different impact of discount factor on different xApps further highlights the distinct objectives defined, with the EE xApp requiring low discount values for optimality. Also, the two xApps showed optimal convergence for different learning rates, namely $a = 10^{-3}$ for the DRM xApp and  $a = 10^{-4}$ for the EE xApp.

Considering the optimal $a$ and $\gamma$ values, there was a noticeable impact of power granularity on long-term reward achievement. Specifically, DRM xApp presented optimal performance for a power step of $P_s = 3$ Watts, leading to an average data-rate improvement of $0.11$ Mbps. This means that following the DRM decisions at a given time slot, the overall system throughput may be increased by $0.11$ Mbps at the next time slot. Considering the continuous operation of DRM xApp, consecutive decisions may lead to considerable data-rate improvements. On the other hand, EE xApp presented performance improvements for higher values of $P_s$ (ranging 9 to 13 Watts), with a $P_s = 11$ Watts showing the highest reward (about 1.3 Mpbs/Watt). Note that, since EE xApp allows for slight throughput degradation (by a factor of $1-\xi$) to achieve power reduction (see (C4') of problem \eqref{problem2}), it is capable of changing more rapidly (i.e. by larger power steps) the power level of the RUs.

\subsection{Comparison of Different Resolution Policies}\label{cmscenarios}

Following the hyper-parameter tuning of each xApp individually, this subsection evaluates the CMF resolution performance under various resolution policies. Since both xApps propose actions concerning the power control of RUs, different resolution policies can be defined to account for handling the direct conflict. To this end, comparative analyses were conducted considering five different resolution policies to illustrate their impact on network KPIs such as system-level data rate and EE. A simulated DRL environment is assumed to be hosted within NDT for purposes of evaluating different candidate actions for power control. To ensure consistency, the NDT environment considered for the proactive action assessment is the same software package used to train both xApps. Under a given resolution policy $i$ at time slot $t$, NDT returns an evaluation score $s_j^i(t)$ for each candidate action $j$, which quantifies the extent to which action $j$ benefits the target KPI defined by policy $i$. Finally, the xApp action with the maximum evaluation score is chosen. 

The resolution policies considered were the following:

\begin{enumerate}
    \item \textbf{Maximum Throughput-based Selection (MaxTS):} At time slot $t$, this policy selects the action of xApp $j$ ($j=1$ for DRM xApp or $j=2$ for EE xApp) that maximizes the system-wide throughput by computing the following score for each candidate action:
    \begin{equation}
        s_j^1(t) = F_1^j(t)
    \end{equation}
    \noindent where $F_1^j(t)$ is given by \eqref{obj1} after applying action $j$.
    
    \item \textbf{Minimum Power-based Selection (MinPS):} This policy resolves the direct conflicts by selecting the power vector that minimizes the total power consumed. Thus, the score of each action $j$ is computed by:
    \begin{equation}
        s_j^2(t) = -\sum_{n=1}^N\sum_{m=1}^M p_{n,m}^j
    \end{equation}
    \noindent where $p_{n,m}^j$ is the power level assigned to RB $m$ of RU $n$ based on the action of xApp $j$.
    
    \item \textbf{Energy Efficiency-based Selection (EES):} This policy selects the action that provides the best ratio of total network throughput divided by the total consumed power \cite{malik2024achieving}.
    \begin{equation}
        s_j^3(t) = F_2^j(t)
    \end{equation}
    \noindent where $F_2^j(t)$ is given by \eqref{obj2} after applying action $j$ and reflects the network EE.
    
    \item \textbf{Throughput Violation-based Selection (TVS):} In this policy, specific SLA requirements are considered for the achieved throughput. We consider an SLA-defined throughput requirement of $c_u$ (in Mbps) for each UE $u$, which reflects that each UE, upon action selection, should experience a throughput of at least $c_u$ Mbps. Without loss of generality, we assumed $c_u = 2$ Mbps for each UE $u$. The evaluation score of this policy quantifies how many UEs exhibit SLA violation by counting the number of UEs (denoted as $C_u$) experiencing a throughput below $c_u$. In case of score ties among xApps (i.e. multiple actions show the same number of SLA violations), the minimum power-consuming action is chosen. The above-mentioned policy returns a score $s_j^4(t)$ for each action $j$, as reflected by the following formula:
    \begin{equation}
        s_j^4(t) = -\sum_{u=1}^{U} C_u^{j}(t) - \frac{1}{1 + e^{-p_{total}^j(t)}}
    \end{equation}
    \noindent where $p_{total}^j(t)$ is the total power consumption (summed across RUs and RBs) after applying action $j$ at time slot $t$, and $C_u^{j}(t)$ is the SLA violation (binary) index of UE $u$ after applying action $j$, which is given by: 
    \begin{equation}
        C_u^{j}(t)=
        \begin{cases}
            1 & \text{, if }  r_u^{j}(t) < c_u \\
            0 & \text{, otherwise}
        \end{cases}
    \end{equation}
    \noindent where $r_u^{j}(t)$ is the experienced throughput (in Mbps) of UE $u$ after applying action $j$. It is evident that this policy scores an action based on the resulting number of SLA violation occurrences (first term) and the total power consumption scaled between [0,1] using the sigmoid function (second term). Given that the number of SLA violations $C_u$ is an integer, the second (sigmoid) term is used to resolve the score ties, ensuring that the contribution of the power consumption on the final score will always be less than one. Under this definition, the second term influences the action selection only when the candidate actions present equality in SLA violations. For instance, if both DRM and EE xApps' actions result in $5$ SLA violations, but they require different total power consumption (e.g., $p_{total}^{DRM}(t)=3 W$ and $p_{total}^{EE}(t)=2 W$, then the returned scores are $s_{DRM}^4(t) = -5-0.95 = -5.95$ and $s_{EE}^4(t) = -5-0.88 = -5.88$. In this example, EE xApp will be finally applied because it produces the highest score (i.e. equal SLA violations compared to DRM xApp, with lower power consumption).

    \item \textbf{EE Violation-based Selection (EEVS):} Similar to TVS policy, this resolution policy considers a UE-level SLA requirement concerning the lower-bounded EE ratio $e_u$ (i.e. a throughput to consumed power ratio). Considering that each UE $u$ should experience at least $e_u = 2$ Mbps/W after RU power allocation, this policy calculated and returns the following score:
    \begin{equation}
        s_j^5(t) = -\sum_{u=1}^{U} E_u^{j}(t) - \frac{1}{1 + e^{-p_{total}^j(t)}}
    \end{equation}
    \noindent where $p_{total}^j(t)$ is the total power consumption (summed across RUs and RBs) after applying action $j$ at time slot $t$, and $E_u^{j}(t)$ is the EE-based SLA violation (binary) index of UE $u$ after applying action $j$, which is given by: 
    \begin{equation}
        E_u^{j}(t)=
        \begin{cases}
            1 & \text{, if }  \frac{r_u^{j}(t)}{p_u^j} < e_u \\
            0 & \text{, otherwise}
        \end{cases}
    \end{equation}
    \noindent where $r_u^{j}(t)$ and $p_u^j$ is the experienced throughput (in Mbps) and the associated RU/RB power level of UE $u$, respectively, after applying action $j$. The EEVS scores quantify how many EE violations occurred and, if score ties between xApps are observed, the least power-consuming action is chosen.     
\end{enumerate}

Notably, the proposed framework is generalizable, meaning that additional resolution policies or modifications of the considered policies could be easily adopted without significant changes in the architectural setup.

\begin{figure}[t]
\centering
\includegraphics[width=0.975\columnwidth]{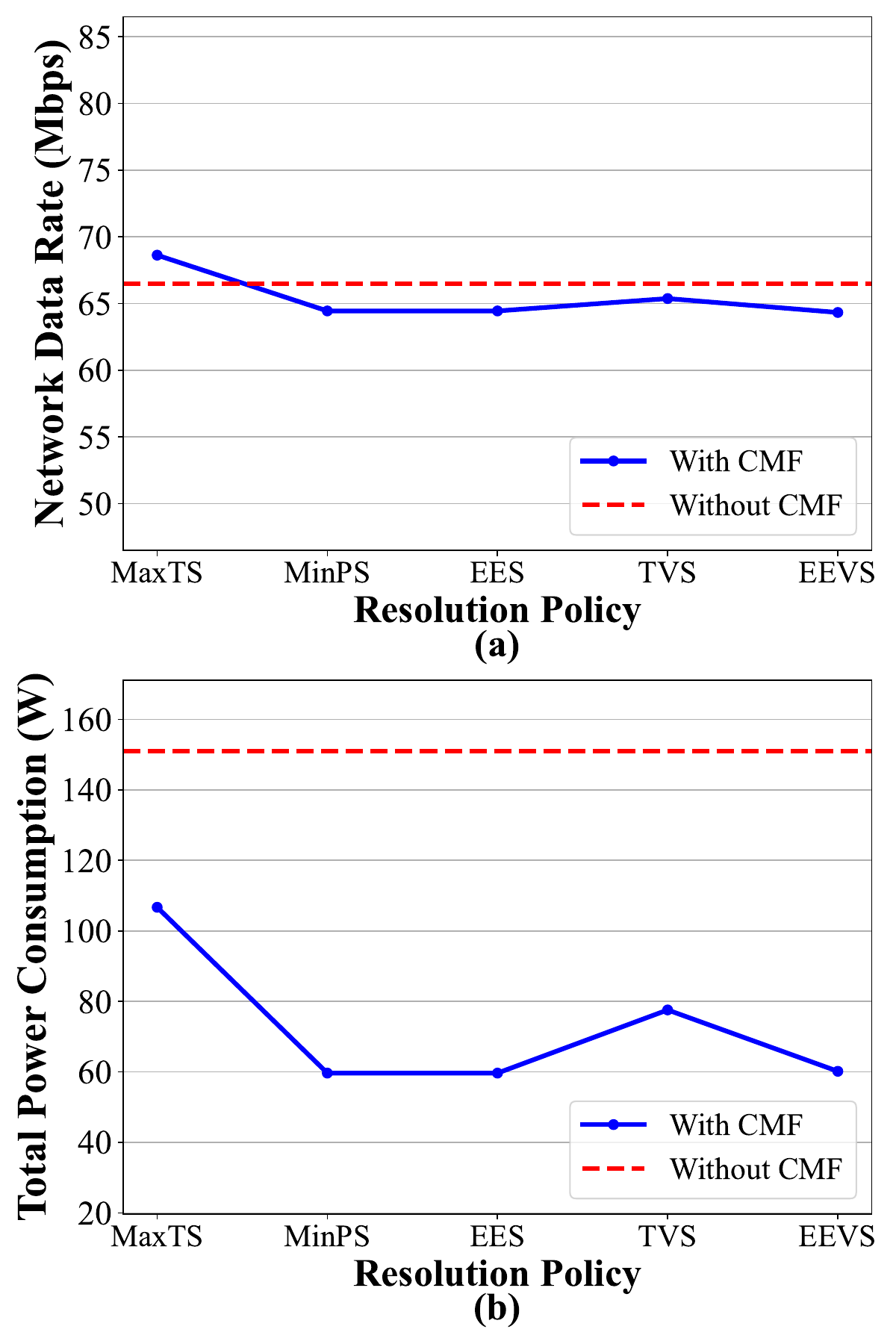}
    \caption{Network Data Rate \textbf{(a)} and Total Power Consumption \textbf{(b)} resulted from different resolution policies (blue curves) after $200$ validation episodes. Red curves reflect the same metrics without the presence of CMF.}
    \label{fig:cm1}
\end{figure}

\begin{figure}[t]
\centering
\includegraphics[width=0.95\columnwidth]{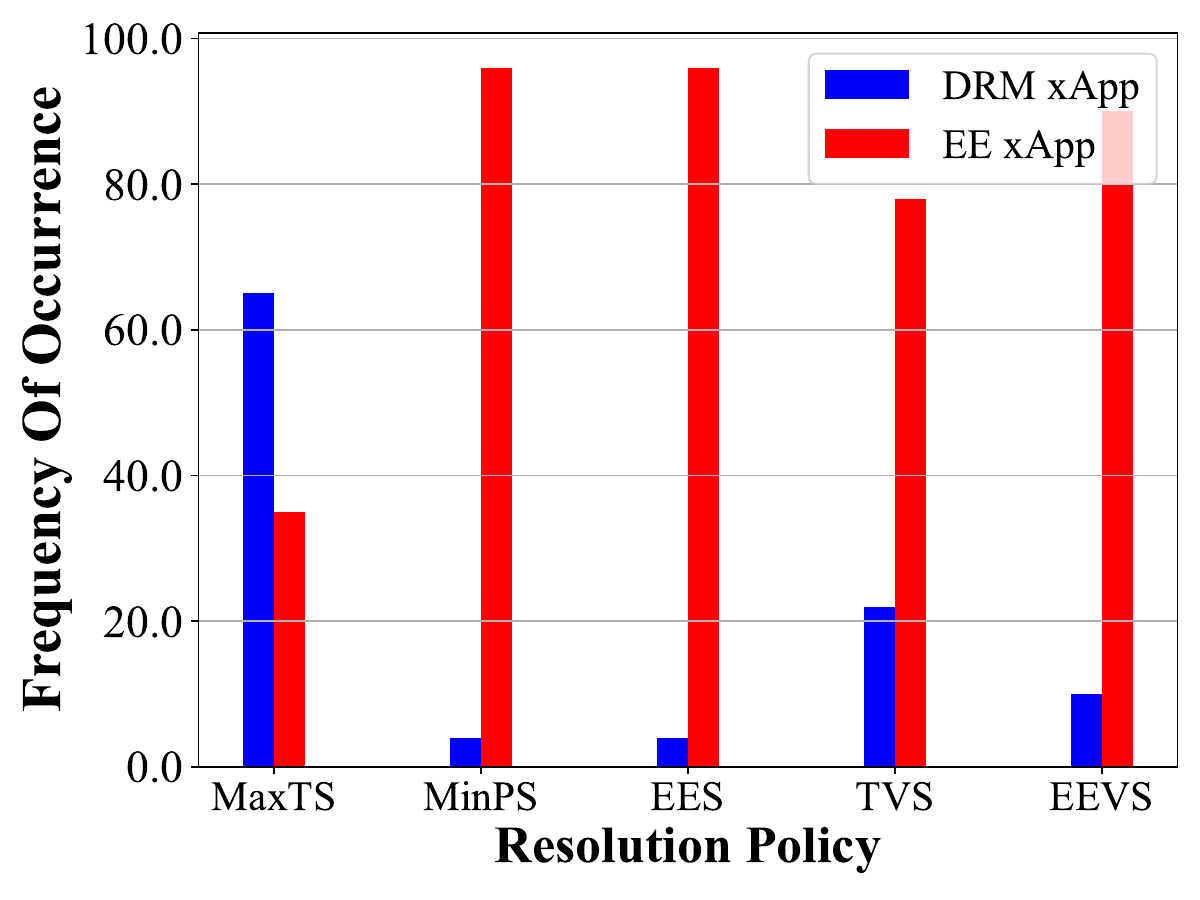}
    \caption{Frequency of occurrence for each xApp's action (DRM and EE) across different resolution policies, showing how many times each xApp's action was finally selected based on the NDT-oriented evaluation scores.}
    \label{fig:cm2}
\end{figure}

To demonstrate the performance of each resolution policy, we conducted $200$ validation episodes, each of which simulating different network instances with randomized initial state (UE position, RU power levels, RB-UE association). In each validation scenario, we run $100$ consecutive steps with both DRM and EE xApps being activated, allowing UEs to randomly move within the network area. In each step, the environment state is acknowledged to both xApps for inference, both producing conflicting power allocation actions that are further detected and resolved by the CMF component. 

After applying the five presented resolution policies hosted in the Conflict Resolver, we measure the Network Data Rate (in Mbps) and the resulted Total Power Consumption (in Watts), averaged across the $200$ episodes and $100$ steps. Both metrics were computed under two conditions: once with the CMF in place (considering a given resolution policy) and once without it. In the CMF-free scenario, the final action is determined based on the timing of the xApp's action arrival, effectively resulting in a random selection between the DRM and EE xApps. This baseline method represents a typical scenario where conflict detection is not performed. Consequently, the most recently received action at the Action Taker is directly applied to the network, overwriting any previously proposed action.

In Fig. \ref{fig:cm1}, we depict the total data rate and the power consumption across the five different resolution policies against CMF-free scenarios. Evidently, regardless of the resolution policy used, CMF inclusion does not significantly benefit the network data rate relative to the CMF-free method. This is attributed to the fact that, at each step, CMF-free scheme leads to the implementation of the action proposed by either DRM or EE xApp, both targeting at the throughput maximization in their objective function (see \eqref{obj1} and \eqref{obj2}). This means that, even without CMF inclusion, both xApps produce optimized decisions with regards to network data rates, further explaining the equal performance between the CMF-based and CMF-free schemes in terms of the overall achieved throughput. Noteworthy, the CMF-based scheme outperforms the CMF-free method only when the MaxTS policy is considered, with the latter constantly selecting the xApp that provides the highest network data rate. Also, the lower data rate observed in all resolution policies compared to the MaxTS is explained by the following key point: MaxTS selects the highest-throughput action, neglecting power consumption and whether UEs are over- or under-satisfied. For instance, an action selected by MaxTS may present extremely high network data rate, simply because some UEs experience extremely high data rate (due to their close proximity from the RUs), while many other UEs are under-satisfied. On the contrary, the rest of the methods either take care of the power consumption (MinPS and EES) or follow an SLA-driven selection (TVS and EEVS), which usually lead lower network-wide data rates.

Importantly, the CMF considerably increases the energy savings of the system, as clearly illustrated in Fig. \ref{fig:cm1}b. Independently of the resolution policy used, CMF inclusion constantly outperforms the CMF-free scheme, highlighting the advantageous usage of CMF for improving system's EE. Following MaxTS, the power consumption is increased compared to the rest of the policies, mainly because MaxTS exclusively selects based on UEs' throughput. The resolution policies that take the power consumption into consideration before selecting the final action (i.e. MinPS, EES and EEVS) showed equally optimal performance in terms of power savings, suggesting that their use can significantly improve EE under the presence of power control-based direct conflicts.

\begin{figure*}[t]
\centering
\includegraphics[trim={0.5cm 0.4cm 0.5cm 0.5cm},clip,width=1.95\columnwidth]{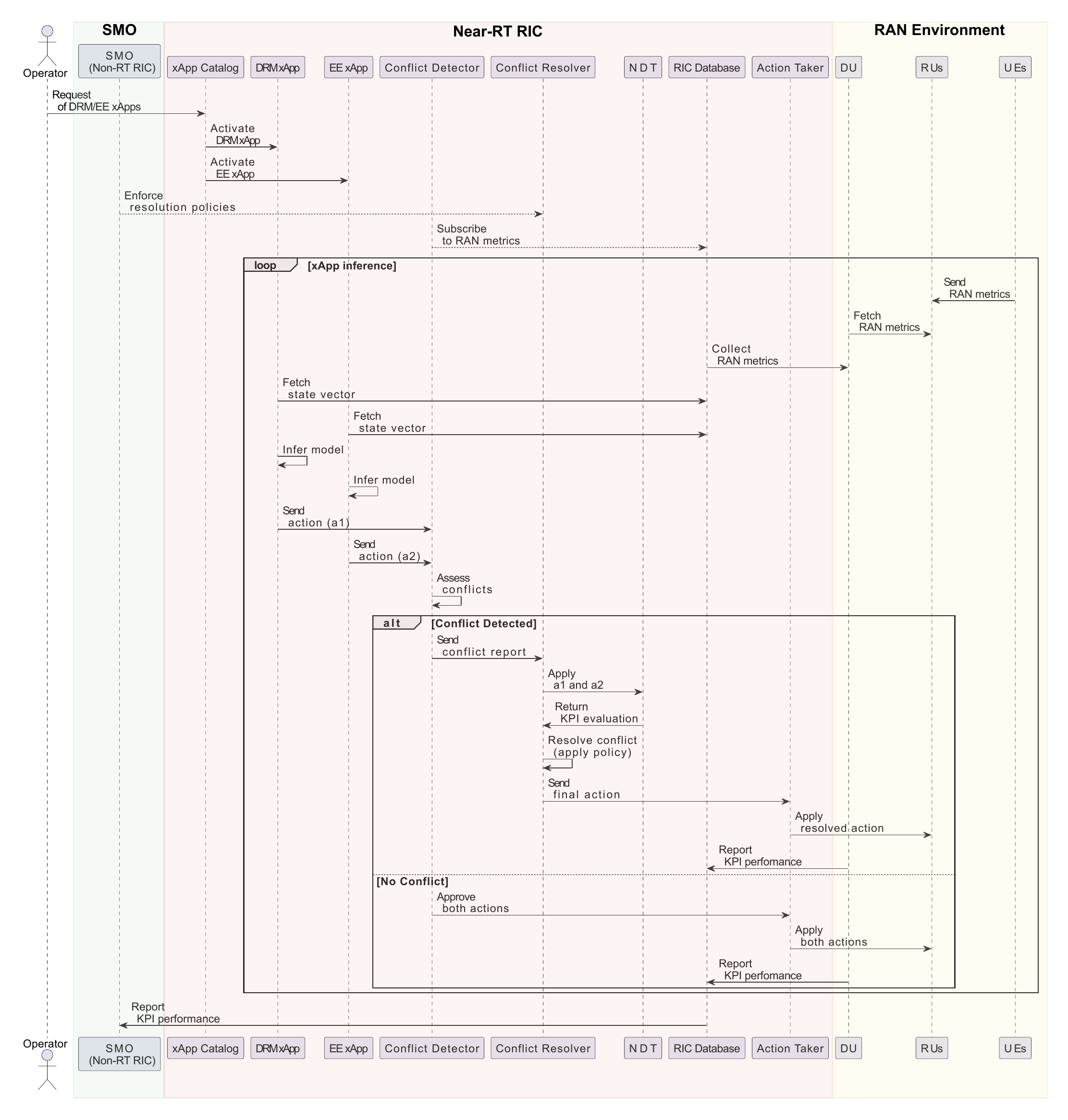}
\caption{Sequence diagram and information flow in the COMIX scheme for multi-xApp inference and conflict management.}
\label{fig:uml}
\end{figure*}

Fig. \ref{fig:cm2} depicts how many times each xApp was finally selected during the validation scenarios. Interestingly, EE xApp is chosen in the vast majority of time across all resolution policies. This is attributed to the fact that EE xApp jointly considers the throughput and the power levels, achieving a balanced rate of both KPIs (data rate and power consumption). Note that DRM xApp is selected only when the objective is to maximize the overall data rate (as in MaxTS policy) and is rarely chosen when power control or SLA violations are considered.

\subsection{End-to-end Workflow for Pre-trained xApps}

To concretely demonstrate the information flow in the COMIX framework, this subsection presents the stepwise workflow to integrate a hierarchical and iterative decision-making under the presence of conflicts \cite{corici2024towards}. The workflow is depicted as sequence diagram in Fig. \ref{fig:uml}, highlighting the interactions among the key components of the O-RAN ecosystem, namely the SMO (Non-RT RIC), Near-RT RIC, DRM and EE xApps, CMF, and the RAN environment (DU, RUs, UEs). The main phases and steps are the following:

\textbf{Phase 1 - Activation of xApps:} Upon operator request towards the Near-RT RIC's xApp Catalog, DRM and EE xApps are activated for purposes of DRL model inference. Once activated, these xApps begin functioning as the primary DRL agents in the workflow.

\textbf{Phase 2 - Policy Enforcement by SMO:}
Then, SMO (Non-RT RIC) enforces the conflict resolution policies onto the Conflict Resolver component. These policies define how potential conflicts should be resolved based on operator-defined objectives, such as prioritizing throughput or energy efficiency (see Section \ref{cmscenarios}). Conflict Detector component subscribes to RIC Database to collect real-time RAN metrics.

\textbf{Phase 3 - Initial State Fetching:}
The DRM and EE xApps retrieve the initial RAN metrics, such as the associated RUs and RBs of the UEs and CQI values, from the RIC Database. This ensures that xApps have the necessary information about the current state of the network to make informed decisions.

\textbf{Phase 4 - xApp Inference and Action Proposals:} Both xApps independently generate power control actions ($a_1$ for DRM and $a_2$ for EE) based on their DRL policies, as defined in \eqref{obj1} and \eqref{obj2}. These proposed actions are then sent to the Conflict Detector for conflict assessment.

\textbf{Phase 5 - Conditional Conflict Resolution:} Conflict Detector assesses the proposed actions as described in Fig. \ref{fig:fig3}. If no conflict is detected, both actions are approved and sent to the RAN environment for direct implementation.
If a conflict is detected, the Conflict Detector informs the Conflict Resolver, which interacts with NDT to obtain the evaluation KPI score of each proposed action. The NDT simulates the impact of both actions on the network, considering metrics such as those presented in Section \ref{cmscenarios}. Based on these evaluations and the enforced policies, the Conflict Resolver selects the optimal action and applies it to the RAN environment through the Action Taker.

\textbf{Phase 6 - Action Application in the RAN:}
The final action, resolved or approved, is applied in the RAN environment (DU or RUs). This adjusts the power levels of the RUs according to the chosen policy, ensuring efficient and conflict-free operation of the network.

\textbf{Phase 7 - KPI Reporting:}
Using RIC Database, the RAN environment continuously reports KPI metrics, such as QoS performance, throughput, and energy consumption, back to the SMO. These reports allow the SMO to monitor network performance and refine its policies over time if needed.

\section{Conclusion}

\subsection{Work Summary}

This work presents COMIX, a comprehensive Conflict Management framework for Multi-Channel Power Control in O-RAN xApps. The framework addresses the critical challenge of managing conflicting objectives among xApps operating on the Near-RT RIC. It also integrates a CMF component for detecting and resolving conflicts and leverages the NDT to simulate the impact of xApp actions before their deployment in the live network. The framework is demonstrated using two DRL-based xApps, namely the DRM and EE xApps. Realistic multi-channel power control scenarios are used for validation process, where various conflict resolution policies are evaluated. The results showcase the effectiveness of power consumption-aware policies in achieving significant energy savings and balanced optimization objectives, outperforming baseline throughput-centric resolution policies and CMF-free methods.

\subsection{Work Extensions}

Future extensions of this work could further enhance the applicability and robustness of the COMIX framework. First, although architectural principles are generalizable, the incorporation of additional xApps introducing direct, indirect and implicit conflicts and targeting diverse objectives, such as latency minimization or fairness, could broaden the scope of the proposed framework. Second, the introduction of adaptive conflict resolution policies, informed by real-time network dynamics and powered by advanced machine learning techniques, could further improve decision-making. Finally, given that conflicts are resolved based on NDT-oriented metrics, elaborating on how NDT could continuously provide accurate representations of the real O-RAN system remains of paramount importance.

\section*{Acknowledgments}
The authors warmly thank Mr Ilias Paralikas for his help in constructing the simulations.

\bibliographystyle{IEEEtran}

\bibliography{mybibfile}

\end{document}